\documentclass[11pt,a4paper]{article}
\usepackage{jheppub}

\usepackage[utf8]{inputenc}

\usepackage{graphicx}
\usepackage{subcaption}
\usepackage[titletoc]{appendix}
\usepackage{xspace}

\usepackage{enumitem}
\setlist{noitemsep}

\usepackage[style=numeric-comp,sorting=none]{biblatex}
\usepackage{lineno}  

\usepackage{comment}

\begin{document}

\noindent
\begin{tabular*}{\linewidth}{lc@{\extracolsep{\fill}}r@{\extracolsep{0pt}}}
 & & HSF-DOC-2020-01 \\
 & & 10.5281/zenodo.4009114 \\
 & & 31 August, 2020 \\ 
 & & \\
\end{tabular*}
\vspace{2.0cm}

\renewcommand{\thefootnote}{\fnsymbol{footnote}}

\title{HL-LHC Computing Review:\\Common Tools and Community Software}

\author{HEP Software Foundation\footnote{\href{mailto:hsf-editorial-secretariat@googlegroups.com}{hsf-editorial-secretariat@googlegroups.com}}:
  Aarrestad, Thea$^{3}$;
Amoroso, Simone$^{5}$;
Atkinson, Markus Julian$^{31}$;
Bendavid, Joshua$^{3}$;
Boccali, Tommaso$^{14}$;
Bocci, Andrea$^{3}$;
Buckley, Andy$^{34}$;
Cacciari, Matteo$^{21}$;
Calafiura, Paolo$^{18}$;
Canal, Philippe$^{7,a}$;
% ...
\bigskip
\par {\footnotesize $^{1}$ High Energy Physics Division, Argonne National Laboratory, Argonne, IL, USA}
\par {\footnotesize $^{2}$ Physics Department, Brookhaven National Laboratory, Upton, NY, USA}
\par {\footnotesize $^{3}$ CERN, Geneva, Switzerland}
\par {\footnotesize $^{4}$ California Institute of Technology, Pasadena, California, CA, USA}
\par {\footnotesize $^{5}$ Deutsches Elektronen-Synchrotron, Hamburg, Germany}
\par {\footnotesize $^{6}$ Frankfurt Institute for Advanced Studies, Johann Wolfgang Goethe-Universität Frankfurt, Frankfurt,Germany}
\par {\footnotesize $^{7}$ Fermi National Accelerator Laboratory, Batavia, IL, USA}
\par {\footnotesize $^{8}$ The Henryk Niewodniczański Institute of Nuclear Physics, Polish Academy of Sciences, ul. Radzikowskiego 152, 31-342 Kraków, Poland}
\par {\footnotesize $^{9}$ Illinois Institute of Technology, Chicago, USA}
\par {\footnotesize $^{10}$ IJCLab, CNRS, Université Paris-Saclay and Université de Paris, Orsay, France}
\par {\footnotesize $^{11}$ INFN Sezione di Bologna, Università di Bologna, Bologna, Italy}
\par {\footnotesize $^{12}$ INFN Sezione di Firenzea, Università di Firenze, Firenze, Italy}
\par {\footnotesize $^{13}$ INFN Sezione di Genova, Genova, Italy}
\par {\footnotesize $^{14}$ INFN Sezione di Pisa, Università di Pisa, Scuola Normale Superiore di Pisa, Pisa, Italy}
\par {\footnotesize $^{15}$ INFN Sezione di Roma Tor Vergata, Roma, Italy}
\par {\footnotesize $^{16}$ Institute of Physics Belgrade, Pregrevica 118, Belgrade, Serbia}
\par {\footnotesize $^{17}$ Institute for High Energy Physics of the National Research Centre Kurchatov Institute, Protvino; Russia}
\par {\footnotesize $^{18}$ Lawrence Berkeley National Laboratory and University of California, Berkeley, CA, USA}
\par {\footnotesize $^{19}$ Laboratoire Leprince-Ringuet, École Polytechnique, CNRS/IN2P3, Université Paris-Saclay, Palaiseau, France}
\par {\footnotesize $^{20}$ Fakultät für Physik, Ludwig-Maximilians-Universität München, München, Germany}
\par {\footnotesize $^{21}$ Laboratoire de Physique Nucléaire et de Hautes Energies (LPNHE), Sorbonne Université, Université Paris Diderot, CNRS/IN2P3, Paris, France}
\par {\footnotesize $^{22}$ Massachusetts Institute of Technology, Cambridge, MA, USA}
\par {\footnotesize $^{23}$ National Taiwan University (NTU), Taipei, Taiwan}
\par {\footnotesize $^{24}$ P.N. Lebedev Physical Institute of the Russian Academy of Sciences, Moscow, Russia}
\par {\footnotesize $^{25}$ Technische Universität Dresden, Dresden, Germany}
\par {\footnotesize $^{26}$ Physics Division, Lawrence Berkeley National Laboratory and University of California, Berkeley CA, USA}
\par {\footnotesize $^{27}$ University of Birmingham, Birmingham, United Kingdom}
\par {\footnotesize $^{28}$ H.H. Wills Physics Laboratory, University of Bristol, Bristol, United Kingdom}
\par {\footnotesize $^{29}$ Department of Physics and Astronomy, University College London, London, United Kingdom}
\par {\footnotesize $^{30}$ University of California, San Diego, La Jolla, CA, USA}
\par {\footnotesize $^{31}$ University of Illinois Urbana-Champaign, Champaign, Illinois, IL, USA}
\par {\footnotesize $^{32}$ University of Cincinnati, Cincinnati, OH, USA}
\par {\footnotesize $^{33}$ IPPP, Durham University, Durham, United Kingdom}
\par {\footnotesize $^{34}$ SUPA - School of Physics and Astronomy, University of Glasgow, Glasgow, United Kingdom}
\par {\footnotesize $^{35}$ Institut für Physik, Humboldt Universität zu Berlin, Berlin, Germany}
\par {\footnotesize $^{36}$ Institut für Astro- und Teilchenphysik, Leopold-Franzens-Universität, Innsbruck, Austria}
\par {\footnotesize $^{37}$ Oliver Lodge Laboratory, University of Liverpool, Liverpool, United Kingdom}
\par {\footnotesize $^{38}$ Université Catholique de Louvain, Belgium}
\par {\footnotesize $^{39}$ Fysiska institutionen, Lunds Universitet, Lund, Sweden}
\par {\footnotesize $^{40}$ School of Physics and Astronomy, University of Manchester, Manchester, United Kingdom}
\par {\footnotesize $^{41}$ Universidad de Oviedo, Instituto Universitario de Ciencias y Tecnologías Espaciales de Asturias (ICTEA), Oviedo, Spain}
\par {\footnotesize $^{42}$ Peking University, Beijing, China}
\par {\footnotesize $^{43}$ Princeton University, Princeton, NJ, USA}
\par {\footnotesize $^{44}$ University of Puerto Rico, Mayaguez, USA}
\par {\footnotesize $^{45}$ Purdue University, West Lafayette, USA}
\par {\footnotesize $^{46}$ University of Washington, Seattle, WA, USA}
\par {\footnotesize $^{47}$ Department of Physics and Astronomy, University of Sheffield, Sheffield, United Kingdom}
\par {\footnotesize $^{48}$ University of Vienna, Austria}
\par {\footnotesize $^{49}$ Department of Physics, University of Warwick, Coventry, United Kingdom}
\bigskip
\par {\footnotesize $^{a}$ Editor}

}

\maketitle


\hypertarget{introduction}{%
\section{Introduction}\label{introduction}}

Common and community software packages, such as ROOT, Geant4 and event
generators have been a key part of the LHC's success so far and
continued development and optimisation will be critical in the future
\cite{stewart_graeme_a_2018_2413005, Ellis:2691414, Albrecht2019}. The
challenges are driven by an ambitious physics programme, notably the LHC
accelerator upgrade to high-luminosity, HL-LHC, and the corresponding
detector upgrades of ATLAS and CMS. The General Purpose Detectors
describe their specific challenges elsewhere; in this document we
address the issues for software that is used in multiple experiments
(usually even more widely than ATLAS and CMS) and maintained by teams of
developers who are either not linked to a particular experiment or who
contribute to common software within the context of their experiment
activity. We also give space to general considerations for future
software and projects that tackle upcoming challenges, no matter who
writes it, which is an area where community convergence on best practice
is extremely useful.

ATLAS and CMS will undergo major detector upgrades and will also
increase their trigger rates for HL-LHC by about a factor of 10; event
complexity will rise, with peak pile-up of 200, far higher than in
Run-2. This places an enormous burden on storage and processing
resources. Current CMOS microprocessor technology is clock speed limited
(due to the failure of Dennard scaling) and, while processors
per-integrated circuit still keeps rising, Moore's Law is expected to
stall during the 2020s. More importantly, the effective runtime related
improvements in computing from CPU servers at sites is likely to be only
about 10\% per year, making the shortfall in processing resources more
severe. As manufacturers struggle to deliver ever-more effective
computing through CPUs the drive to different architectures intensifies,
with GPUs becoming commonplace and increasing interest in even more
specialised architectures, such as TPUs, IPUs and developments that make
FPGA devices more user friendly \cite{acm10.1145/3282307}.
These pose huge challenges for our
community as the programming models and APIs vary widely here and
possible lock-in to a particular vendor's devices is a significant
problem for code sustainability and preservation. Huge work has been
done already to adapt to this changing landscape, e.g. multi-threading
software and GPU codes have been used in production by some experiments
for years now~\cite{calafiura2018hep, albrecht2018hepexec}.
In other areas migration is still ongoing. In yet others it has
not even started. Generic heterogeneous programming models have
existed for some time, and new ones are arising, but there is, as yet,
no clear winner, in-part because the performance obtained can be highly
application-dependent. HEP itself will not drive the success of one
model or another, so even if the community converged on a preferred
model, its long term success would not be assured. The C++ standard
itself is likely to lag for many years behind what is required by us in
the area of heterogeneous or even distributed computing. Further,
experiment frameworks (and with knock-on effects to systems like
workload management) will need to adapt to sites that provide
heterogeneous resources. How to do this, and make effective use of
different heterogeneous resources across different sites, remains far
from settled.

For storage systems (see the DOMA and WLCG documents) the pressure on
software is to store as much physics information in as few bytes as
possible, but also to be able to read at very high rates to deliver data
from modern storage technologies into the processing hardware. This
requires critical developments in the storage formats and libraries used
in HEP, e.g. developments like RNTuple for ROOT I/O is likely to be of
great importance for the community \cite{ROOT-2020-HL-LHC}. The likelihood of finding an
off-the-shelf compact and efficient storage format for HEP data is
remote, so investment in smart software to support our PB sized science
data is simply cost effective. Particularly for analysis, which is
usually I/O bound, we have an end-to-end problem from storage
technology, through the software layers, to processing resources that
may well span multiple nodes. Other workflows, which are less dependent
on I/O rates will, nevertheless, have to be adapted to using remote data
where the I/O layer must optimise data transfers and hide latency, e.g.
taking advantage of XRootD's single column streaming ability \cite{xrootd}.

In this increasingly complex environment in which to write software,
there are important problems where sharing information at the community
level is far more efficient. Providing a high level of support in the
build environment for developers, sharing knowledge about how to
measure, and then improve, performance (especially on multiple different
architectures) and sharing best practice for code development can have a
large integrated benefit \cite{couturier2017hep}. This requires improved training and the
development of a curriculum for all developer levels. In the field, such
initiatives are often warmly received, but real support is needed for
those who can put this into practice, also ensuring that their work in
training contributes to their career development and a long term future
in the field \cite{foundation2018hep}. HEP software stacks are already deep and wide and building
these consistently and coherently is also an area where knowledge can be
shared. Support is needed for multiple architectural targets and
ensuring the correct runtime in heterogeneous environments.

This brings up the important question of validation and the need to
improve the security of physics results, which is even more complicated
on heterogeneous platforms, when exact binary compatibility often cannot
be assured. Currently almost every experiment and project has its own
infrastructure for this.

Once software is built, it needs to be shipped worldwide so that the
production workflows can run. CernVM-FS was a huge leap forward for
software distribution and has even been widely adopted outside HEP 
\cite{Blomer_2011, 7310920}.
However, new challenges arise, with container based payloads, scaling
issues and disconnected supercomputer sites. So maintenance and
development needs to be undertaken to support and extend this key
supporting infrastructure for software.

Finally, over the multi-decade lifetimes of HEP experiments, we need to
preserve both the core and analysis software so that results can be
confirmed and updated as the field moves on. There are many exciting
developments based around CernVM-FS
\cite{10.1007/978-3-319-67630-2_52, serverless-cvmfs}, containers and things like analysis
description languages \cite{Ref10}, but these are not yet at the stage of being
settled nor integrated into our day-to-day workflows.

In the rest of this document the main issues associated with the key
parts of the software workflow in high-energy physics are presented,
focusing on those that dominate current resource consumption: physics
event generation, detector simulation, reconstruction and analysis.

\hypertarget{physics-event-generators}{%
\section{Physics Event Generators}\label{physics-event-generators}}

\hypertarget{introduction-1}{%
\subsection{Introduction}\label{introduction-1}}

Physics event generators are essential in HEP. All of the LHC scientific
results, both precision measurements or searches for new physics, depend
significantly on the comparison of experimental measurements to
theoretical predictions computed using generator software.

Using Monte Carlo (MC) techniques, generators allow both the generation
of unweighted events for experimental studies of differential
distributions and the prediction of total cross sections. The
large-scale event generation campaigns of the LHC experiments have
significant computational costs, mainly in terms of CPU resources. The
limited size of simulated samples is a source of major uncertainty in
many analyses and is therefore a limiting factor on the potential
physics output of the LHC programme. This situation will get
significantly worse for 
HL-LHC. The fraction of the CPU resources used for event generation today is approximately 5-15\%. 
As is the case for the other big
consumers of CPU (detector simulation and reconstruction), speedups in generator software are
needed to address the overall resource problem expected at the HL-LHC,
compounded because more accurate predictions, requiring more complex
calculations will be
needed (e.g. beyond NLO or with higher jet multiplicities). Many other issues, both technical and
non-technical exist, e.g. funding, training, careers for those working in this
area~\cite{Alves:2017she,Gen18}.

A HSF Working Group (WG) on generators~\cite{Gen19} was set up at the
beginning of 2019. The main focus of the WG so far has been on gaining a
better understanding of the current situation, and identifying and
prioritising the areas where computing costs can be reduced. In
particular, the WG has been active in analysing the ATLAS and CMS
compute budgets in detail, in profiling MG5\_aMC~\cite{Alw14}, 
Sherpa~\cite{Bot19} and Powheg~\cite{Fri07}, in discussing the possible sharing
of common parton-level samples by ATLAS and CMS, and in reviewing and
supporting the efforts for porting generators to modern architectures
(e.g., MG5\_aMC to GPUs). This last activity is particularly important,
as it has become increasingly clear that being able to run
compute-intensive WLCG software workloads on GPUs would allow the
exploitation of modern GPU-based supercomputers at High Performance
Computing (HPC) centres, and generators look like a natural candidate
for this, as they are smaller code bases without complex dependencies.

This section gives an overview of the many technical and non-technical
challenges in the generator area and of the work that can be done to
address them. This is finally condensed into a list of a few
high-priority items, for the next 18 months. A more detailed version of this
contribution, including a more complete list of references, is
uploaded to arXiv and will be submitted for publication~\cite{Gen20}.

\hypertarget{collaboration-challenges}{%
\subsection{Collaborative Challenges}\label{collaboration-challenges}}

\subsubsection{Generator Software Landscape} 

The landscape of generator software is extremely varied. Different
generators are used for different processes. Generating a sample also
involves choices of precision (e.g. Leading Order, LO, or
Next-to-Leading-Order, NLO), hadronisation and Parton Shower (PS)
models, underlying event tunes, prescriptions for matching/merging and
simulating particle decays, and other input parameters, chiefly among
them the parton density functions, for which different interfaces exist.
Various combinations of software libraries are thus possible, often
written by different authors and frequently many years old. For a
given process the LHC experiments often use different software packages
and settings from each other, and a single experiment can generate
events using more than one choice. Many different packages and
configurations may therefore need to be worked on to get cumulative CPU
cost reductions. The large number of external packages also complicates
their long-term maintenance and integration in the experiments' software
and workflows, sometimes leading to job failures and computing
inefficiencies. Other packages are also absolutely critical for the
whole generator community and must be maintained, even if their CPU cost
is relatively low (LHAPDF, Rivet, HepMC, etc.).

\subsubsection{Skills and Profiles}

A very diverse combination of skills and profiles are needed for generator software.
Theorists
(who create fundamental physics models, and design, develop and optimise
most generator code), experimentalists working in research (who request
different samples) and in computing (who implement, monitor and
account execution of workflows on computing resources), software
engineers and system performance experts. This is a richness and
opportunity, as some technical problems are best addressed by people
with specific skills, but it also poses some challenges:

\paragraph{Training challenges.} Theorists and experimentalists often
lack formal training in software development and optimisation. Software
engineers and experimentalists are often not experts in the theoretical
physics models implemented in MC codes.

\paragraph{Communication challenges.} It is difficult to find a shared
terminology and set of concepts to understand one another: notions and
practices that are taken for granted in one domain may be obscure for
others. As an example, there are many articles about the physics in
generators, but software engineers need papers describing the software
modules and overall control and data flow.

\paragraph{Career challenges.} Those working in the development, optimisation or
execution of generator software provide essential contributions to the
success of the (HL-)LHC physics programme and it is critical that they
get the right recognition and motivation. However, theorists get
recognition on published papers, and may not be motivated to work on
software optimisations that are not ``theoretical'' enough to advance
their careers. Generator support tasks in the experiments may also not
be enough to secure jobs or funding for experimentalists pursuing a career
in research.

\paragraph{Mismatch in usage patterns and optimisation focus.} The way
generators are built and used by their authors is often different from
the way in which they are deployed and integrated by the experiments in
their software frameworks and computing infrastructure. The goals and
metrics of software optimisation work may also differ. Theorists are
mainly interested in calculating cross sections and focus on minimising
the phase space integration time for a given statistical precision. The
LHC experiments run large scale productions of unweighted event
generation, and mainly need to maximise the throughput of events
generated per unit time on a given node.

\paragraph{Programming languages.} Attracting collaborators with a computer
science background to work on generators, especially students, may also
be complicated by the fact that critical components of some generator
packages are written in Fortran, which is rarely used in industry and
less popular among developers than other programming languages. Some of
the generators also do not use industry standard version control
systems, making it harder to contribute code.

\hypertarget{technical-challenges-software-and-computing}{%
\subsection{Technical Challenges}\label{technical-challenges-software-and-computing}}

The event generation workflow presents several challenges and
opportunities for improvement.

\subsubsection{Inefficiency in Unweighted Event Generation}

\paragraph{Phase space sampling inefficiency.} Efficient sampling is the most
critical ingredient for efficient unweighted event generation. Many
generic algorithms exist (e.g. VEGAS~\cite{Lep80}, BASES / SPRING~\cite{Kaw86}, 
MINT~\cite{Nas07}, FOAM~\cite{Jad03}), as well as others
developed specifically for a given generator (e.g. MadEvent~\cite{Mal03},
itself based on a modified version of VEGAS, in MG5\_aMC, or COMIX~\cite{Gle08} 
in Sherpa). In general, the larger the dimensionality of the
phase space, the lower the unweighting efficiency that can be achieved:
in W+jets, for instance, the Sherpa efficiency is 30\% for W+0 jets and
0.08\% for W+3jets~\cite{Gao20}. This is an area where research is very
active, and should be actively encouraged, as significant cost
reductions in WLCG compute budgets could be achieved. Improvements in
this area start from physics-motivated approaches based on the
knowledge of phase space peaks and are complemented by
machine learning (ML) algorithmic methods~\cite{Ben17,Bot20,Gao20,Kli18}.

\paragraph{Merging inefficiency.} Merging prescriptions (e.g. MLM~\cite{Man02},
CKKW-L~\cite{Lon02} at LO and FxFx~\cite{Fre12}, MEPS@NLO~\cite{Hoe14} at
NLO) imply the rejection of some events, to avoid double counting
between events produced with n\textsubscript{jets}+1 matrix elements and
with n\textsubscript{jets} MEs plus parton showers. The resulting
inefficiencies can be relatively high, depending on the process, but they
are unavoidable in the algorithmic strategy used by the underlying
physics modeling~\cite{Alw08}. However, a method like shower-kt MLM can
reduce the merging inefficiency of MLM~\cite{Alw09}.

\paragraph{Filtering inefficiency.} An additional large source of inefficiency
is due to the way the experiments simulate some processes, where they
generate large inclusive event samples, which are then filtered on
final-state criteria to decide which events are passed on to detector
simulation and reconstruction (e.g. CMS simulations of specific
$\Lambda_{B}$ decays have a 0.01\% efficiency and ATLAS B-hadron
filtering has \textasciitilde10\% efficiency for V+jets). This
inefficiency could be reduced by developing filtering tools within the
generators themselves, designed for compatibility with the requirements
of the experiments. Filtering is an area where LHCb has a lot of
experience and already obtained significant speedups through various
techniques. The speed of colour reconnection algorithms is a limiting
factor for simulating rare hadron decays in LHCb.

\paragraph{Sample sharing and reweighting.} In addition to removing
inefficiencies, other ways could be explored to make maximal use of the
CPU spent for generation by reusing samples for more than one purpose.
Sharing parton-level samples between ATLAS and CMS is being discussed
for some physics analyses. Event reweighting is already commonly used
(e.g. for new physics searches and some systematic uncertainty
calculations) and could be explored further, though this may require
more theoretical work for samples involving merging or NLO matching~\cite{Mat16}.

\paragraph{Negative weights.} Matching prescriptions, like MC@NLO, are required
in (N)NLO calculations to avoid double counting between (N)NLO matrix
elements and parton showers, leading to the appearance of events with
negative weights. This causes a large inefficiency, as larger event
samples must be generated and passed through the experiment simulation,
reconstruction and analysis codes, increasing the compute and storage
requirements. For a fraction $r$ of negative weight events, the number of
events to generate increases by a factor $1/(1-2r)^2$: for instance,
with $r=25\%$ (which may be regarded as a worst-case scenario in top quark
pair production~\cite{Fre20}), one needs to generate 4 times as many
events. Negative weights can instead be almost completely avoided, by
design, in another popular matching prescription, POWHEG; however, this
is only available for a limited number of processes and describes the
relevant physics to a different degree of precision than MC@NLO 
(see~\cite{Fre20} for a more in-depth discussion). Progress in this area can
only be achieved with physics knowledge: for instance, a new approach for
MC@NLO-type matching with reduced negative weights has recently been
proposed~\cite{Fre20} and some recent papers show how to lessen the impact of
negative weights in a secondary step~\cite{andersen2020positive,
nachman2020neural}. It should be noted that negative weights can also happen at
LO because of not-positive-definite parton density function sets and
interference terms, e.g. in effective field theory calculations.

\subsubsection{Accounting and Profiling}

While progress has been made to better understand which areas of
generator software have the highest computational cost, more detailed
accounting of the experiment workloads and profiling of the main
generator software packages would help to further refine R\&D
priorities.

\paragraph{Accounting of CPU budgets for generators in ATLAS/CMS.} Thanks to a
lot of effort from the generator teams in both experiments, a lot of
insight into the settings used to support each experiment's physics
programme was gained within the WG, and it is now clear that the
fraction of CPU that ATLAS spends for event generation is somewhat
higher than that in CMS. More detailed analysis of the different
strategies is ongoing. These figures had to be
harvested from logs and production system databases, which was
particularly difficult for CMS, requiring significant person hours.
It is important to establish better
mechanisms to collect this information, to allow for an easy comparison
between different experiments.

\paragraph{Profiling of the generator software setups used for production.}
Another area where the WG has been active is the definition and
profiling of standard generator setups, reproducing those used in
production. Detailed profiling could also be useful to assess the CPU
cost of external parton distribution function libraries~\cite{Kon20}, or
the memory requirements of the software (which may motivate a move to
multithreading or multiprocessing approaches).

\subsubsection{Software Modernisation}

More generally, as is the case in other areas of HEP, some R\&D on
generator software is certainly be needed to modernise it and make it
more efficient, or even port it to more modern computing 
architectures~\cite{Bau13,Alves:2017she}:

\paragraph{Data parallelism, GPUs and vectorisation.} The data flow of an MC
generator, where the same (matrix element) function is computed
repeatedly at many phase space points, lends itself naturally to the
data parallel approaches found in CPU vectorised code and in GPU compute
kernels. Porting and optimising generators on GPUs is essential to be
able to use modern GPU-based HPCs. The work done in this direction in
the past on MG5\_aMC, that never reached production quality, is now
being reinvigorated by the WG, in collaboration with the MG5\_aMC team, and
represents one of the main R\&D priorities of the WG. This work is
presently focusing on Nvidia CUDA, but abstraction libraries will also
be investigated. GPUs may also be relevant to ML-based sampling
algorithms and to the pseudo-random number generation libraries used in
all MC generators.

\paragraph{Task parallelism, multithreading and multiprocessing.} Generators are
generally executed as single threaded software units. In most cases,
this is not a problem, as the memory footprint of unweighted event
generation is small and usually fits within the 2 GB per core available
on WLCG nodes. However, there are cases where memory is higher than 2 GB
(e.g. DY+jets using Sherpa); this leads to inefficiencies as some
processor cores remain unused, which could be avoided using
multithreading approaches. The fact that some generators are not even
thread safe may also be a problem, for instance to embed them in
multi-threaded event processing frameworks, such as that of CMS.
Multi-processing approaches may also be useful to speed up the
integration and optimisation step for complex high-dimensional final
states. In particular, a lot of work has been done to implement
MPI-based multi-processing workflows for generators in recent years. For
instance, the scaling of LO-based generation of merged many-jet samples
has been successfully tested and benchmarked on HPC architectures using
both Alpgen~\cite{Chi17} and Sherpa~\cite{Hoe19}; in the latter case, new
event formats based on HDF5, replacing LHEF, have also been instrumental
in the success of these tests. MPI integration has also been completed
for MG5\_aMC~\cite{Mat18}. It should also be noted that, even if HPCs
offer extremely high-speed inter-node connectivity, it is perfectly ok
for WLCG workflows to use these systems as clusters of unrelated nodes.

\paragraph{Generic code optimisations.} A speedup of generators may also be
achievable by more generic optimisations, not involving concurrency. It
should be studied, for instance, if data caching~\cite{Kon20} or
different compilers and build strategies may lead to any improvements.

\hypertarget{physics-challenges}{%
\subsection{Physics Challenges}\label{physics-challenges}}

In addition to software issues, important physics questions should also
be addressed about more accurate theoretical predictions (above all NNLO
QCD calculations, but also electroweak corrections) and their potential
impact on the computational cost of event generators at HL-LHC. Some
specific NNLO calculations are already available and used today by the
LHC experiments in their data analysis. With a view to HL-LHC, however,
some open questions remain to be answered, in particular:

\paragraph{NNLO: status of theoretical physics research.} The first question is,
for which processes NNLO precision will be available at the time of the
HL-LHC? For example, when would NNLO be expected for $2 \rightarrow 3$ or
even higher multiplicity final states? And for lower multiplicity final
states, where differential NNLO predictions exist but the generation of
unweighted NNLO+PS events is not yet possible. It is important to
clarify the theoretical and more practical challenges in these
calculations, and the corresponding computational strategies and impact
on CPU time needs.

\paragraph{NNLO: experimental requirements for data analysis at HL-LHC.} The
second question is, for which final states unweighted event generation
with NNLO precision would actually be required? and how many events
would be needed? One should also ask if reweighting LO event samples to
NNLO would not be an acceptable cheaper alternative to address the
experiment's needs.

\paragraph{Size of unweighted event samples required for experimental analysis
at HL-LHC.} Another question, unrelated to NNLO, is in which regions of
phase space the number of unweighted events must be strictly
proportional to the luminosity? For example, in the low $p_\text{T}$ regions of W
boson production it is probably impossible to keep up with the data, due
to the huge cross section. Alternative techniques should be
investigated, to avoid the generation of huge event samples.

\hypertarget{detector-simulation}{%
\section{Detector Simulation}\label{detector-simulation}}

\hypertarget{introduction-2}{%
\subsection{Introduction}\label{introduction-2}}

In HEP, data analysis, theory model evaluation and detector design
choices rely on detailed detector simulation. Since the start of the
first run the LHC experiments have produced, reconstructed, stored,
transferred and analysed hundreds of billions of simulated events. This
effort required a major fraction of WLCG's computing 
resources~\cite{ALICE-TDR-12, ATLAS-TDR-17, CMS-TDR-7, LHCb-TDR-11, WLCG-CM-Update, 
ALICE-TDR-019, LHCb-TDR-018, ATLAS-LHCC-2019-02, CMS-LHCC-2019-09}

The increase in delivered luminosity in future LHC runs, in particular
with the HL-LHC, will create unique research opportunities by collecting
an order of magnitude more data. At the same time, the demand for
detector simulation will grow accordingly so that
statistical uncertainties are kept as low as possible. However, the expected
computing resources will not suffice if current detector simulation is
used; the availability of simulated samples of sufficient size
would soon become a major limiting factor as far as new physics
discoveries, for example through precision measurements, are concerned.
Development of faster simulation, therefore, is of crucial importance
and different techniques need to be explored under the assumption that
they will provide a sufficient gain in time performance with a
negligible or acceptable loss in physics accuracy.

The Geant4 simulation toolkit has been the de facto standard for HEP
experiment simulation for physics studies over the last two decades~\cite{Geant4-2003, Geant4-2006, Geant4-2016}.
Designed in the late 1990s to cope with the simulation challenges of the
LHC era, Geant4 introduced flexible physics modelling, exploiting layers
of virtuality. This fostered progress in physics by comparison of
competing models on the same energy ranges, and allowed complementary models
to be combined to cover large energy ranges. The simulation applications
of the LHC experiments make use of this toolkit and, as such, most of
the LHC physics program relies on it. It has delivered a physics
precision that allows the experiments to perform precision analysis of
the collected data and to produce new results pushing the boundaries of
our understanding in HEP. At the same time, the code of the toolkit is
becoming old, with some parts written almost thirty years ago, making it
more and more difficult to run efficiently on modern computing
architectures. Any improvement in the critical elements of Geant4 can be
leveraged by the applications that use it. 
Improvements of physics models need to continue to take place to prevent
systematic uncertainties from simulation from becoming a dominant
factor, however, their exact influence on the precision of physics
analysis is not always easy to quantify. On the other hand, the overall
CPU performance of the current code will not improve drastically just by
relying on modern hardware or better compilers. 

Taking all those elements into account, three main development paths have emerged to be
undertaken simultaneously and these have started to be pursued. However,
effort in any one of them is far from sufficient to exploit their
individual potential. 

Firstly, it is necessary to continue to modernise
and refactor the simulation code by using more optimal, modern
programming techniques. 


Secondly, different fast simulation techniques, where tracking of
individual particles in given detectors is replaced by some sort of
parameterisation and precision is traded to achieve higher event
throughput.

Finally, the use of compute
accelerators (like GPUs or FPGAs) is the third emerging path that needs
to be undertaken and that requires intense R\&D effort.


We will discuss
more in detail these three development axes in the following sections, 
outlining the developments identified in and since the
Community White Paper~\cite{Albrecht2019, SimFoundation2018hep}. 

\hypertarget{geant4-rd}{%
\subsection{Geant4 R\&D}\label{geant4-rd}}

The Geant4 particle transport toolkit contains advanced models for
electromagnetic and hadronic physics processes, as well as the
geometrical description of the detectors. The toolkit design principles
were to allow flexibility in future extensions of its own code, as well
as choices for particular applications. It is clear that this
flexibility comes at the price of a performance penalty. It was observed that
writing more compact and more specialised code can lead in certain
places to up to several tens of percent speed-up~\cite{GeantV-2019, GeantV-2020}. The
technical improvements to explore here consist of avoiding too many
virtual layers and improving the instruction and data locality, leading
to better cache use.

Going even further, as far as specialisation is
concerned, studies suggest that implementing compact, self-contained
transport libraries with a minimal set of physics models (for instance
only electromagnetic interaction for selected particles) with predefined
scoring (to simulate, for instance, the response of an EM calorimeter)
can also bring speed-ups to the complete application~\cite{GeantV-2019, GeantV-2020}.

As
far as architectural modifications are concerned, the design of the
Geant4 track-status machinery prevents the introduction of fine-grained
track parallelism. It is still not clear whether this kind of
parallelism can bring any substantial speed-up, due to the overheads;
however, making the Geant4 kernel and physics processes stateless would
certainly provide the necessary starting points for further R\&D in that
direction.

Most of these investigations have started and it is hoped to
see some first results on a year long time-scale. More effort, however, will
be required to integrate any eventual changes in the production code in
a timely manner.

The GeantV Vector prototype has allowed the assessment
of the achievable speedup using a novel, vectorised approach to particle
transport. The demonstrator of a full electromagnetic shower in
realistic (CMS) geometry has been implemented and the comparison to a
similar Geant4 application has been performed. The conclusion of that
work showed that the main factors in the speedup seem to include better
cache use and tighter code, while the vectorisation impact was much
smaller than hoped for. This unimpressive impact is due to a set of
factors, including the fraction of the current algorithms that could be
vectorised, unresolved challenges for the gather/scatter and tail
handling (to keep the number of events in flight within bound) needed
for large number of shapes, and of course Amdahl's Law applied to
vectorisation where some bottleneck (for example geometry navigation)
could not be tackled without a major additional long term effort~\cite{GeantV-2019, GeantV-2020}.

At the same time libraries developed in the context of the
GeantV R\&D, e.g. VecGeom, have been successfully integrated in Geant4
and ROOT benefiting the whole HEP software community~\cite{CMS01}. The
cost-over-benefit of more vectorisation of the code has not been deemed
worth the necessary investment also due to the speculation that a
complete rewrite of the physics base could become necessary to fully
exploit it. As a result investing in the optimisation and modernisation
of the Geant4 code has gained even more relevance.

\hypertarget{experiments-applications-and-optimisation-of-the-use-of-geant4}{%
\subsection{Experiment Applications and Optimised Use of
Geant4}\label{experiments-applications-and-optimisation-of-the-use-of-geant4}}

All LHC experiments have dedicated simulation frameworks making use of
the Geant4 toolkit for modelling physics processes in their detectors. Every experiment
configures and uses what is provided by Geant4 for their specific needs.
Effort has been spent by each experiment since the publication of the
Community White Paper to benchmark their 
simulation code and to maximally
exploit the options provided by the Geant4 toolkit to optimise the
performance of their applications~\cite{ATLAS-G4OPT}. All of the various handles provided
are being continuously explored and adopted when deemed useful: range
and physics cuts have been reviewed and customised by all experiments,
shower libraries adopted as baseline in the forward region by ATLAS and
CMS, stepping granularity optimised, magnetic field caching
investigated, neutron Russian roulette used for dedicated simulations.
All of these can have major impacts on simulation time and sharing
knowledge and experience between experiments can be of high benefit even
if each experiment has to find its optimal solution.

ATLAS, CMS and LHCb have integrated the Geant4 event-based
multi-threading features within their own multi-threaded frameworks,
harmonising their different architectural choices with the Geant4
implementation~\cite{CMS02, LHCb01, ATLAS01}, and have used, or are planning to
use, them in their production environment to gain access to a larger set
of distributed resources. ATLAS has been running their multi-process
simulation framework on HPCs systems for production in the
last years~\cite{ATLAS02, Benjamin:2696330} 
while CMS has pioneered the use of their multi-threaded
simulation on available resources including HPCs~\cite{CMS03}. Both
approaches have allowed the exploitation of additional computing
resources by enabling access to lower memory systems.

\hypertarget{fast-simulations}{%
\subsection{Fast Simulations}\label{fast-simulations}}

Fast simulation techniques consist of replacing the
tracking of individual particles through the detector (or part thereof),
including all the physics processes they would undergo, by a
parameterisation, where the detector response is produced directly as a
function of the incoming particle type and energy.

An example of such a
parameterisation was implemented many years ago in the context of the H1
experiment~\cite{H1Gflash}. This parameterisation, available within the
Geant4 toolkit under the name of the GFLASH library, became the starting
idea for several custom implementations, specific for other
calorimeters. Those were interfaced to experiments' simulation
applications through a hook in the Geant4 toolkit. The LHC
experiments implemented, for example, dedicated parametrised response
libraries for some of their calorimeters. Recently, an effort has been
invested in Geant4 to generalise the parameterisation formulae, and to
implement automatic procedures of tuning the parameters for specific
detector geometries. This kind of `classical' fast simulation, which
describes the shower shape using some complex mathematical
functions, with several parameters, will remain an important tool;
however, it is also clear that their overall precision for different
energy values, especially for highly granular and complex calorimeters
will always be relatively low. 

Recent developments of deep
learning-based techniques have opened up an exciting possibility of
replacing those `classical' parameterisations by trained neural networks
that would reproduce the detector response. This approach consists of
training generative models, such as Generative Adversarial Networks
(GAN), Variational Auto-Encoders (VAE) or Autoregressive Generative
Networks on the `images' of particle 
showers~\cite{Paganini_2018, ML001-GAN, aishik_ghosh_2019_3599705, GAN-ATLAS}. The energy
deposition values in the calorimeter cells are considered as `pixels' of
the images that the network is supposed to reproduce. The first studies
and prototypes have shown very promising results, however, the
generalisation of the developed models (for different calorimeters, with
different particles incident at different angles with different
energies), still requires further effort. Detailed physics validation of
those tools and understanding their capability to reproduce
fluctuations is a prerequisite towards production quality fast
simulation libraries. This work is already ongoing, but will require,
over the next few years, more effort to be invested, combining
physics and Machine Learning expertise. Given that the training of the
deep learning algorithms usually requires using sufficiently large MC
samples produced using standard techniques, the development of these novel
tools does not necessarily remove the need to speed up Geant4.

It is worth pointing out that different fast simulation techniques have
already been extensively used for LHC simulation campaigns. Effort has
been spent in the experiments frameworks to combine as much as possible
fast and detailed implementations. In particular, ATLAS and CMS are using
shower libraries in production for the modelling of
calorimeters in the forward region, in combination with the detailed
simulation of the rest of the sub-detectors. In LHCb the re-use of part
of the underlying events for specific signals of interest has been
established in production of samples when appropriate, with particular
care paid to ensure keeping under control bias on statistical
uncertainties~\cite{LHCb02}. Applicability of this technique in ATLAS and CMS could be
explored for specific cases. Similarly, the re-use of simulated or real
events to mimic the background to hard-scatter events is also being exploited,
where additional challenges exist concerning storage and I/O due to the
handling of the large minimum bias samples needed to model the
additional interactions.

All LHC experiments have explored the use of multi-objective regression
and generative adversarial networks (GANs)~\cite{GAN-ATLAS, GAN-LHCb, aishik_ghosh_2019_3599705}, they are now in the process of investigating the use of
fast simulations for other types of detectors than calorimeters, e.g.
Cerenkov based systems~\cite{Lamarr} and Time Projection Chambers~\cite{ALICE-TPC}. While implementations of fast simulations of given
sub-detectors is specific to their technology and experimental
environment and as such the responsibility of each LHC experiment, an
area of potential common investigation is frameworks for fast tuning and
validation.

The use of fully parametric response of experimental setups at the level
of reconstructed objects (tracks and clusters) for rapid evaluation of
physics reach is also exploited by the experiments with generic
frameworks for fast simulation of typical collider detectors being used~\cite{bib-Delphes, PGS4}. 
The advantage of such frameworks is that they allow
simultaneous studies for different experiments as well as their
potential use by the theory community. Experiment specific fully
parameterised simulations providing reconstructed objects compatible
with the experiment's analysis frameworks for systematic verifications
have also been emerging~\cite{Lamarr} .

It is reasonable to expect that the full variety of simulation options,
from `smeared reconstructed quantities' to parameterisation for specific
detectors to detailed Geant4, will need to be exploited by the
experiments for different tasks. Seamless ways of providing them in a
transparent and integrated way in the experiments' simulation frameworks
should continue to be pursued with Geant4 support.

\hypertarget{technical-challenges-software-and-computing-1}{%
\subsection{Technical Challenges}\label{technical-challenges-software-and-computing-1}}

The hardware landscape has always set directions as far as
software evolution is concerned. The recent adaptation of the
simulation code to multithreading (MT) turned out to be relatively
straightforward, but still took several years to adopt and validate.
Geant4 can now run in MT mode efficiently, distributing the events
between threads and sharing resources, like the geometry description
or physics data, and thus reduce the memory footprint of a simulation
while maintaining its throughput performance. The efficient utilisation
of Single Instruction Multiple Data (SIMD) architectures, on the other
hand turned out to be more challenging. Not only was a limited part of
the code vectorisable but also, as the already mentioned GeantV R\&D
has shown, overheads related to gathering the data for vector processing
presented a significant challenges (tail handling, work balancing across
threads, etc.) to efficiently (development time wise and run time wise)
profit from it~\cite{GeantV-2019, GeantV-2020}.

The most recent revolution in computing hardware is the
use of Graphics Processing Units (GPUs) for general purpose computing. Several
attempts have been made to port specific simulation code to GPUs. A few
of them turned out to be successful, leading to factors of several
hundred or a thousand speedup~\cite{GPU-MPEXS-DNA, Hybrid-Gate, Opticks}; however, they were always
limited to a very specific simulation problem, far from what is
addressed by a general HEP simulation. This application of GPUs has been
very successful in restricted domains like medical physics simulation,
neutron transport~\cite{SHIFT01} or optical photon transport~\cite{CHEP-Juno}. 
In those applications, the type of particles considered
is very limited (often just one type, with no secondary particle
production), the set of physics processes is reduced and the geometry is
often extremely simple compared to the LHC detectors. A natural
extrapolation of those approaches to a general HEP simulation is very
difficult to imagine, because the stochasticity of the simulated events
would immediately lead to divergences as far as different GPU threads
are concerned, not to mention, simply the feasibility of porting the
whole simulation to GPU code, for which specific expertise would be
needed. 

On the other hand, running only very small pieces of the
simulation application on GPUs does not seem to be efficient either, as
gathering data and transferring it from the host to the device and back again
may strongly limit any possible performance gain, similar to what was
seen with the GeantV vectorisation R\&D. The most plausible approaches
seem therefore to lead in the direction of specialised libraries (like
those described above in the context of speeding up Geant4 simulation on
the CPUs) that would perform the complete simulation of specific
sub-detectors (for instance EM calorimeter, for specific incoming
particles or Cherenkov-based detectors for the optical processes) on the
device. Such libraries could be the right compromise between the
complexity of the algorithm that needs to be ported and
the overall time that is now saved in
by the full simulation on the CPU.

An investigation of
these directions is starting now, but it will certainly require
considerable effort to be invested before being able to judge
the real impact. The first steps that are currently being undertaken
consist of implementing, on GPUs, prototypes based on VecGeom geometry
and navigation capabilities~\cite{Wenzel_2017}. If those prototypes turn out to be
successful, the next challenge will consist of adding tracking and
eventually a limited subset of physics models. While Geant4 is a toolkit
capable of addressing different modelling and simulation problems and
contains many features and capabilities allowing for user access to
detailed information for a large variety of use cases and scientific
communities, this GPU development might turn into specialised transport
modules, stripped of some features which would be expensive to implement
or support efficiently on GPUs.

Another interesting avenue consists of
exploring the application of some vendor libraries (like Nvidia Optix).
Those libraries, originally developed for ray tracing, have several
similarities with general particle transport and, if applied
successfully, could lead to major speed-ups. All these efforts certainly
require a major investment and new developers, expert in those
technologies, to join the R\&D work.

\hypertarget{other-activities}{%
\subsection{Other Activities}\label{other-activities}}

Common digitisation efforts would be desirable among experiments, with
advanced high-performance generic examples, which experiments could use
as a basis to develop their own code. Nevertheless the large variety of
detector technologies used reduces its
applicability. Digitisation is not yet a limiting factor, in terms of
CPU requirements, so developments in this area have not been a priority.

Simulated samples are often processed through the reconstruction as real
data. As such in some cases they require the emulation of hardware
triggers. Hardware triggers are based on very specific custom devices
and a general approach does not seem very feasible, even if some
parametrisation could be generalised.

\hypertarget{outlook}{%
\subsection{Outlook}\label{outlook}}

It is important to realise that there is a considerable
risk that simulation becomes a major limiting factor as far as new
physics discoveries are concerned if no serious effort is invested in
R\&D. Concerted effort of different experts in physics,
Machine Learning and GPUs is needed. There are too many unknowns to focus on only
one direction, so a lot of different prototyping is needed. Development of Geant4,
while working on fast simulation algorithms and
doing extensive R\&D on leveraging compute accelerators is required. While the R\&D
activities are taking place, one can not forget that it is still
necessary to support the existing code and make sure that sufficient
funding and staffing is provided for maintenance and development of
physics algorithms, as well as for adapting the code to any updated CPU
hardware, operating systems and new compilers.

It is also important to stress once again the need for a continuous
integration activity of new runtime performance
improvements and solutions into the Geant4 code base and experiments
applications to profit from any improvement as quickly as possible. An
effort needs to be invested into making it possible to evolve it to meet
the HL-LHC needs without compromising the production quality of the
Geant4 code used by the experiments. We need to ensure that new
developments resulting from the R\&D programs can be tested with
realistic prototypes and, if successful, then integrated, validated, and
deployed in a timely fashion in Geant4 and adopted by the experiments.
The strategy adopted in the successful integration in Geant4 of VecGeom
libraries developed in the context of the GeantV R\&D can provide a
working example of how to proceed to provide some incremental
improvements to existing applications.

No single solution appears at the moment to provide by itself the
processing gains required by HL-LHC, nevertheless if a novel approach
emerges that could do so it should be pursued and carefully evaluated in
terms of gains against the disruption of a complete re-implementation of
how events are simulated.

\hypertarget{reconstruction-and-software-triggers}{%
\section{Reconstruction and Software
Triggers}\label{reconstruction-and-software-triggers}}

Software trigger and event reconstruction techniques in HEP face a
number of new challenges in the next decade. Advances in facilities and
future experiments bring a dramatic increase in physics reach, as well
as increased event complexity and rates.

At the HL-LHC, the central challenge for high-level triggers (e.g.
software triggers) and object reconstruction is to maintain excellent
efficiency and resolution in the face of high pileup values, especially
at low transverse momenta. Detector upgrades, such as increases in
channel density, high precision timing and improved detector layouts are
essential to overcome these problems. The subsequent increase of event
complexity at the HL-LHC also requires the development of software
algorithms that can process events with a similar cost per-event to
Run-2 and Run-3. At the same time, algorithmic approaches need to
effectively take advantage of evolutions in computing technologies,
including increased SIMD capabilities, increased core counts in CPUs,
and heterogeneous hardware.

This section focuses on the challenges identified in the Community White
Paper~\cite{Alves:2017she,albrecht2018hepexec,albrecht2018hep} and the
development of solutions since then. It also includes
mentions of open source software efforts that have developed within or
outside LHC collaborations that could be adapted by the LHC community to
improve trigger and reconstruction algorithms.

\hypertarget{the-evolution-of-triggers-and-real-time-analysis}{%
\subsection{Evolution of Triggers and Real-Time Analysis}\label{the-evolution-of-triggers-and-real-time-analysis}}

Trigger systems are evolving to be more capable, both in their ability
to select a wider range of events of interest for the physics program of
their experiment, and their ability to stream a larger rate of events
for further processing. The event rates that will be processed by
experiments at the HL-LHC will increase by up to a factor 10 with
respect to Run-3, owing to upgraded trigger systems and expanded physics
programs. ATLAS and CMS plan to maintain a two-tiered trigger system,
where a hardware trigger makes use of coarse event information to reduce
the event rate to 10x over the current capability, up to 1 MHz \cite{ATLAS-TDR-29,collaboration:2714892}. The high level trigger system, implemented in software, selects
up to 100 kHz of events to be saved in full for subsequent
analysis\footnote{LHCb \cite{Aaij:2019uij} and ALICE \cite{Buncic:2011297} will both
  stream the full collision rate to real-time or quasi-real-time software
  trigger systems.}.

Experiments have also been working towards minimising the differences
between trigger (online) and offline software for reconstruction and
calibration, so that more refined physics objects can be obtained
directly within the HLT farm for a more efficient event selection. This
is also in-line with enhancing the experiment's capabilities for
real-time data analysis of events accepted by the hardware trigger
system, driven by use cases where the physics potential improves when
analysing more events than can be written out in full with
traditional data processing.

Implementations of real-time analysis systems, where raw data is
processed into its final form as close as possible to the detector (e.g.
in the high-level trigger farm), are in use within several experiments.
These approaches remove the detector data that typically makes up the
raw data kept for offline reconstruction, and keep only a limited
set of analysis objects reconstructed within the high level trigger or
keepig only the parts of an event associated with the
signal candidates, reducing the required disk space.
The experiments are focusing on the technological developments that make it
possible to do this with acceptable reduction of the analysis
sensitivity and with minimal biases. The HSF Reconstruction and Software Trigger
group has been encouraging cross-talk between LHC experiments and beyond
on real-time analysis, as these kinds of workflows increase the physics output for selected
physics cases without adding significant overhead to the current
resources.

Active research topics also include compression and custom data formats;
toolkits for real-time detector calibration and validation which will
enable full offline analysis chains to be ported into real-time; and
frameworks which will enable non-expert offline analysts to design and
deploy real-time analyses without compromising data taking quality.
Use cases for real-time analysis techniques have expanded during the
last years of Run 2 and their use appears likely to
grow already during Run 3 data taking for all experiments owing to
improvements in the HLT software and farms. 
Further ideas include retaining high-rate, but very reduced data, from selected regions and
sub-detectors, even up to the 40 MHz bunch crossing rate~\cite{hannes_sakulin_2019_3598769}.

\hypertarget{challenges-and-improvements-foreseen-in-event-reconstruction}{%
\subsection{Challenges and Improvements Foreseen in
Reconstruction}\label{challenges-and-improvements-foreseen-in-event-reconstruction}}

Processing and reducing raw data into analysis-level formats, event
reconstruction, is a major component of offline computing resource
needs, and in light of the previous section it is relevant for online
resources as well. This is an essential step towards precision
reconstruction, identification and measurement of physics-objects at
HL-LHC.

Algorithmic areas of particular importance for HL-LHC experiments are
charged-particle trajectory reconstruction (\textit{tracking}), including
hardware triggers based on tracking information which may seed later
software trigger and reconstruction algorithms; jet reconstruction,
including the use of high-granularity calorimetry; and the use of
precision timing detectors.

\hypertarget{tracking-in-high-pile-up-environments}{%
\subsubsection{Tracking in High Pile-up
Environments}\label{tracking-in-high-pile-up-environments}}

The CPU needed for event reconstruction in Runs 2 and 3 is
dominated by charged particle reconstruction (tracking). This is still a
focus for HL-LHC triggering and event reconstruction, especially when
the need to efficiently reconstruct low transverse momentum
particles is considered.

The huge increase in the number of charged particles, and hence the
combinatorial load for tracking algorithms, at future colliders will put
great strain on compute resources. To minimise the memory footprint,
tracking software needs to be thread-safe and to support multi-threaded
execution per core. At the same time, the software has to be efficient
and accurate to meet the physics requirements.

Since the CWP, experiments have made progress towards improving software
tracking efficiency in HL-LHC simulation, e.g. \cite{ATL-PHYS-PUB-2019-041}. A
number of collaboration and community initiatives have been focusing on
tracking, targeting both offline and online, and these are listed below.

The \textbf{ACTS} project \cite{Ai:2019kze,Gumpert_2017} is an attempt to encapsulate
the current ATLAS track reconstruction software into an
experiment-independent and framework-independent tracking software,
designed to fully exploit modern computing architectures. It builds on
the tracking experience already obtained at the LHC, and targets the
HL-LHC as well as future hadron colliders. ACTS provides a set of track
reconstruction tools designed for parallel architectures, with a
particular emphasis on thread-safety and concurrent event
reconstruction. ACTS has active collaborators from other HEP
experiments, such as FASER, and provides support to Belle-II, EIC and CEPC.

The aim of the \textbf{mkFit} \cite{cerati2019speeding} project is to speed up Kalman filter (KF)
tracking algorithms using highly parallel architectures and to deliver
track building (and possibly fitting) software for the HL-LHC. Recent
activities focused on delivering a production quality software setup to
perform track building in the context of LHC Runs 2 and 3. The
implementation relies on single-precision floating point mathematics and is
available for multicore CPUs. It achieves significant gains in compute
performance from the use of vector instructions, extending to
AVX-512, based on Matriplex library (a part of the mkFit project).

\textbf{Exa.TrkX}~\cite{Ju:2020xty} is a cross-experiment collaboration of data scientists
and computational physicists from ATLAS, CMS and DUNE. It develops
production-quality deep neural network models, in particular Graph
Neural Networks, for charged particle tracking on diverse detectors
employing next-generation computing architectures such as HPCs. It is
also exploring distributed training and optimisation of Graph Neural
Networks (GNN) on HPCs, and the deployment of GNNs with microsecond
latencies on Level-1 trigger systems.

\hypertarget{addition-of-timing-information-in-reconstruction}{%
\subsubsection{Adding Timing Information to
Reconstruction}\label{addition-of-timing-information-in-reconstruction}}

Physics performance in very high pileup environments, such as the HL-LHC
or FCC-hh, may also benefit from adding timing information to the
reconstruction. This allows the mitigation of the effects of pile-up by
exploiting the time-separation of collision products~\cite{Collaboration:2623663,CMS:2667167}

Experiment communities have been working on timing detector
reconstruction and object identification techniques in complex
environments. Since the CWP, initial tracking and vertexing algorithms
that include timing information have been developed and incorporated
into experimental software stacks \cite{LindseyGreyMIPCTD}\footnote{Tracking algorithms where events overlap in time
within time slices are employed by the CBM experiment~\cite{Akishina:2015ghv,AkishinaThesis}. The use of a cellular automaton makes the
  algorithm less dependent on the specific detector geometry, and there
  may be the possibility of cross-talk with LHC experiments that are
  investigating 4D reconstruction.}.

\hypertarget{enhanced-data-quality-and-software-monitoring-for-trigger-and-reconstruction}{%
\subsection{Enhanced Data Quality and Software Monitoring}\label{enhanced-data-quality-and-software-monitoring-for-trigger-and-reconstruction}}

At HL-LHC, the development, automation, and deployment of extended and
efficient monitoring tools for software trigger and event reconstruction
algorithms will be crucial for the success of the experimental physics
programme.

HEP experiments have extensive continuous integration systems, including
code regression checks that have enhanced the monitoring procedures for
software development in recent years. They also have automated
procedures to check trigger rates as well as the performance of the low-
and high-level physics objects in the data. 

Since the CWP, experiments have started making limited use of machine
learning algorithms for anomaly detection \cite{Adinolfi:2298467,CMSMonitoring}.

\hypertarget{general-reconstruction-software-improvements-vectorisation}{%
\subsubsection{General Reconstruction Software Improvements:
Vectorisation}\label{general-reconstruction-software-improvements-vectorization}}

Improving the ability of HEP developed toolkits to use vector units on
commodity hardware will bring speedups to applications running on both
current and future hardware. The goal for work in this area is to evolve
current toolkit and algorithm implementations, and develop best programming
techniques to better use the SIMD capabilities of current and future
computing architectures. Since the CWP, algorithm development projects
have demonstrated success in increasing the use of vector
units~\cite{cerati2019speeding, LHCB-FIGURE-2019-002}. In addition, HEP has increasingly benefited from industry
developed software, including machine learning toolkits (e.g.,
TensorFlow), that typically make excellent use of vector units.
Challenges remain in this area: for example, full exploitation
must also account for many generations of hardware that experiments must
exploit on the grid. In addition, to realise full performance gains, a
large fraction of the application must be improved to use SIMD processing
capabilities (partly due to turbo-boost capabilities of processors).

\hypertarget{use-of-machine-learning-for-software-trigger-and-reconstruction-algorithms}{%
\subsubsection{Use of Machine Learning}\label{use-of-machine-learning-for-software-trigger-and-reconstruction-algorithms}}

It may be desirable, or even necessary, to deploy new algorithms that
include advanced machine learning techniques to manage the
increase in event complexity without increasing per-event
reconstruction resource needs.

Work is already ongoing in the collaborations to evolve or rewrite
existing toolkits and algorithms focused on their physics and technical
performance at high event complexity (e.g. high pileup at HL-LHC), and
efforts in this area have expanded. Cross-collaboration developments are
focusing on the availability of ML algorithms for experimental software,
especially in persistifying models and running inferences in production
and in enhancing the
training capacities of collaborations by submitting jobs to the grid and
to facilities offering large CPU/GPU resources~\cite{LWTNN, ONNX}. One can also foresee
that HEP will increase the benefit from techniques and cross-talk from
industry.

\hypertarget{algorithms-and-data-structures}{%
\subsubsection{Algorithms and Data Structures}\label{algorithms-and-data-structures}}

Computing platforms are generally evolving towards having more cores
or to adopt different
architectural models (GPUs, FPGAs) to increase processing capability~\cite{acm10.1145/3282307}.
The goal for HEP is to improve the throughput of
software trigger and event reconstruction applications, so current
event models, toolkits and algorithm implementations have to evolve to
efficiently exploit these opportunities. The first algorithms that should be
targeted are those which are most time consuming.

Since the CWP, HEP now has a number of algorithmic projects that are
designed for, or have successfully adapted to, hardware accelerators,
most notably to NVIDIA GPUs using CUDA. Two projects targeting Run-3 and
with prospects for use in Run-4 are Patatrack (from CMS) \cite{andrea_bocci_2019_3598824} and
Allen (from LHCb) \cite{Aaij:2019zbu}.

\textbf{Patatrack} is a CMS initiative aimed at using heterogeneous
computing for charged particle reconstruction. Within the CMS
codebase (CMSSW), this project has demonstrated that physics
reconstruction code can be written to leverage heterogeneous
architectures, like GPUs, and achieve a significant speed-up, while
reducing the overall cost and power consumption of a system. An example
is the CMS Pixel local reconstruction and the track and vertex
reconstruction: these algorithms can run on an NVIDIA T4 GPU with the
same performance as 52-56 Xeon cores, at a fraction of the cost and
power consumption of an equivalent CPU-only system; further improvements
may come pairing one or more high end GPUs with a low power ARM system.

\textbf{Allen} is a data processing framework for GPUs, as well as a
specific implementation of a first-level trigger for LHCb Run-3
data-taking. Allen is optimised for sustaining the rate required by
real-time processing in LHCb, but can be used as a more general GPU data
processing system as it includes a scheduler and a memory manager that can be
integrated into Gaudi~\cite{gaudi_paper}. This permits the extension and implementation of
Allen within other experimental software using the same underlying
framework.

ATLAS is also investigating GPU solutions for reconstruction software
and its acceleration at HL-LHC, in particular in terms of highly
parallel algorithms \cite{attila_krasznahorkay_2019_3599103} and to enable the use of machine learning
algorithms in object reconstruction and identification. 

The experience acquired with these projects shows that GPUs can be used
for increasing the throughput in cases of large-size (ALICE)
\cite{david_rohr_2019_3599418} and small-size events per second (LHCb), and to speed up
specific parts of the data processing (CMS Patatrack).

It is clear that the technology is improving rapidly in this area. The
adoption of portability toolkits is needed to avoid vendor lock-in
and/or the need to evolve algorithms by hand to adapt to each new type
of computing architecture, as well as to increase the sustainability of
the codebase avoiding multiple implementations of the same algorithm.


The \textbf{HLS4ML} project \cite{Duarte:2018ite,Summers:2020xiy,DiGuglielmo:2020eqx} targets the implementation of
machine learning algorithms on FPGA for low-latency applications useful
for e.g. Level-1 triggers. The aim of the HLS4ML project is the
translation of trained ML models into FPGA firmware. HLS4ML also extends
to the production of coprocessor kernels for FPGAs for longer latency
applications, and can be used as a tool to design AI-powered ASICs. The
\textbf{SONIC} application \cite{Duarte:2019fta} is being developed in parallel, with
the goal of facilitating and accelerating the inference of deep neural
networks for triggering, reconstruction, and analysis, by providing
software to use heterogeneous computing resources as a service targeting
next-generation facilities at the energy and intensity frontier (HL-LHC,
LBNF).

\hypertarget{trigger-and-reconstruction-software-sharing}{%
\subsection{Trigger and Reconstruction Software
Sharing}\label{trigger-and-reconstruction-software-sharing}}

Nearly all software solutions presented in this chapter are tailored to
a specific experiment, and would require additional work to be adapted
to different environments. Nevertheless, it is still useful to maintain
open communication channels to discuss design choices and compare
performance assessments of different solutions to guide future software
and reconstruction design choices.

The use of open-source elements of the LHC software stacks for trigger
and reconstruction by smaller experiments, especially in case of common
experimental design\footnote{An example of software sharing happens in
  the case of the FASER experiment, whose offline software
  (\href{https://gitlab.cern.ch/faser/calypso}{{https://gitlab.cern.ch/faser/calypso}})
  is based on a derivative of the ATLAS Athena open-source software, which in turn uses the Gaudi framework. FASER
  also shares tracking detector components with ATLAS, facilitating the
  adoption of software relying on similar detector descriptions.}, is
still worthwhile This benefits smaller experiments
and increases the return on investment for LHC
experiment software. For this reason, we also encourage the addition of
common open software projects to the \href{https://projectescape.eu/services/open-source-scientific-software-and-service-repository}{ESCAPE software catalogue}.

\hypertarget{data-analysis}{%
\section{Data Analysis}\label{data-analysis}}

\hypertarget{key-analysis-computing-challenges-at-the-hl-lhc}{%
\subsection{Key Analysis Computing Challenges at the
HL-LHC}\label{key-analysis-computing-challenges-at-the-hl-lhc}}

Examination of collision data is, in essence, the primary objective of the
experiment collaborations, coming at the end of the data
preparation, simulation and reconstruction chain. The stage of analysis,
starting in most cases from a standard data format generically referred
to as ``Analysis Object Data'' (AOD) containing reconstructed physics
object data, and producing physics results as the end product, poses
unique computing challenges. The scope of analysis data processing is
broad, encompassing the production of derived data formats as well as
end stage analysis actions, such as producing histograms from those
intermediate datasets and visualising the final results. In the
following, attention is focused on the computing challenges related to
analysis for the ATLAS and CMS experiments.

Today, ROOT format AOD files and derived datasets take up the lion's
share of disk resources, filling approximately 80\% of the total data
volume for both ATLAS and CMS \cite{Ref1,Ref2}. To serve precision analyses
that require large event statistics experiments will increase the
recorded event rate by an order of magnitude, compared with a projected
\textasciitilde10\% annual growth of storage resources. The large pile-up in HL-LHC
collisions will compound the storage challenge by inflating the data
size per event. Reducing the storage footprint of data analysis is
therefore of paramount importance.

The effective CPU needs for end-stage analysis payloads are typically
orders of magnitude lower than those in simulation and reconstruction.
When scaled up to O(100) analyses per year, each processing the input
data dozens of times a year, the total CPU consumption at the HL-LHC is
projected to be around 10\% of total experimental computing \cite{Ref1,Ref2}.
Central production of analysis data formats may account for another
10-30\%. It is worth noting at this point that analysis data access
patterns tend to be more chaotic than preceding stages, which can
increase the effective resource needs by large factors.

More significant than the global storage and computational costs is the
job turnaround time, closely tied to the ``time to insight'', which is a
strong limitation on the speed of analysis publication. One significant
constraint is the need for full coverage of the input data, which is
inextricably linked to weaknesses in book-keeping tools that make it
difficult, if not impossible, to make meaningful studies on subsets of
the data. This full coverage requirement is clearly limited by computing
infrastructure load and downtime. Another challenge is the
multiplication of computing resource demands from the assessment of
systematic uncertainties, which involves processing many alternate
configurations and input data, inflating both the CPU load and storage
footprint. Improvement of community standards for metadata handling and
uncertainty handling will be very important for HL-LHC.

A last concern is that, while simulation and reconstruction code is
mostly optimised and developed under greater scrutiny, analysis code is
often produced with emphasis on quick results and less emphasis on code
quality or performance. Over time, this may result in inefficient,
under-documented code that exacerbates the disk and CPU shortfall and
poses a major difficulty when software is re-purposed for a new analysis.
Add to this a growing array of alternative toolkits from the data
science community, in particular for machine learning, and it is clear
that the entropy of the analysis code ecosystem has become a major
challenge in itself.

The following sections expand on the major challenges for analysis
software, including those identified above. An overview is given of
ongoing work on common tools that alleviate these problems, followed by
an outlook on R\&D prospects that would more significantly transform the
HL-LHC analysis model for major resource savings.

\hypertarget{analysis-processing-and-workflows-current-problems-and-solutions}{%
\subsection{Analysis Processing and Workflows: Current Problems and
Solutions}\label{analysis-processing-and-workflows-current-problems-and-solutions}}

To make data analysis tractable, AOD data are typically reduced both by
saving only relevant information for each event, which may require
transformations (object calibration, computation of derived variables)
beyond simply discarding information, and by skimming out only events
that are interesting. In post-reconstruction data formats, lossy
compression is also in use and being optimised. Coordination of the data
reduction phase is a key point for the organisation of data analysis
processing at HL-LHC.

Two approaches have been followed during LHC Run 2 by ATLAS and CMS:
ATLAS analysis trains \cite{Ref4} and the CMS mini-AOD, a highly reduced and
standardised event content. Analysis trains place software payloads
tailored for specific analyses (carriages) into a periodically scheduled
centralised execution, amortising data staging costs and sharing some
common processing steps \cite{Ref3}. Standardised event contents are instead
common reductions that satisfy the needs of the bulk of analyses,
avoiding duplication of commonly selected events in multiple datasets.
Neither of these approaches provides exhaustive coverage of all the analysis use
cases for the experiments and alternative data formats are needed for the
small but significant (10-20\%) fraction of analyses requiring custom
reconstruction that may be CPU-intensive. These exceptions to the rule
are expected to remain at the HL-LHC.

It is worth considering the viability of the arguably more flexible
option to have analysis trains run over the reconstruction output at the
HL-LHC. Extrapolating using a simple model, ATLAS finds that they would
produce 35 PB/year of data AOD and 213 PB/year of MC \cite{Ref1}. In
comparison, a reduced DAOD\_PHYSLITE format (10 kB/event) made from the AOD would
result in 0.5 PB/year for data and 2 PB/year for MC. Production of this
reduced format using a Data Carousel model \cite{Ref1} allows the much
larger AOD to reside on tape, providing significant savings in disk
storage. Experience from CMS shows that further reduction (down to
\textasciitilde2 kB/event for the nano-AOD) could be possible while
still supporting a majority of physics analyses \cite{Ref5}. File size
reductions also permit multiple copies to be held on disk for more
efficient parallel processing. It is clear that as many analyses as
possible need to be fed into this standardised analysis event-content
pipeline if analysis computing resources are to be kept under control.
Nevertheless, analysis trains or any similar means of synchronising data
access could be applied to this highly reduced format. It is worth
noting the LHCb analysis model already in Run 3 has to forego the luxury
of reprocessing raw data. This constraint from sheer event rate is not
without merit as it greatly simplifies the data processing and analysis
models and removes one tier of offline data products.

Careful attention needs to be paid to the assessment of systematic
uncertainties, particularly with regard to event content
standardisation, to avoid creating many derived outputs that differ only
minimally. Analysis models that do not require all event information to
be present in a single file, instead leveraging ROOT ``friend trees'' or
columnar storage, may be a way to reduce this duplication both in the
case of uncertainty calculations and reconstruction augmentations, but
will require development of robust strategies for keeping the
augmentations in sync with the core events. It is noted that the ROOT
data format is quantifiably very efficient for the processing model in
HEP \cite{Ref16}, and any competitors would have to achieve a similar level
of performance.

Access to metadata, defined as cross-event information, is clearly a
weak point in the handling of the multiple analysis datasets originating
from the LHC. Metadata may include information such as book-keeping of
processed events, detector conditions and data quality, a
posteriori measured scale factors, and software versioning. This
information is often scattered in multiple locations such as:
in-file metadata, remote database resources, shared-area text or ROOT files,
TWiki pages or even in e-mail. The lack of proper metadata integration
and a coherent interface leaves analyses exposed to any problems in data
processing. This in turn results in data completeness demands that
simply will not work at the HL-LHC, where the huge datasets all but
guarantee that some fraction of the data will be unavailable at any
given point in time.

At the HL-LHC, as energy and luminosity conditions are quasi stable
across the years, datasets from multiple years with different conditions
will need to be analysed in a coherent way. The analysis software will
then need to be able to fetch and use the proper metadata automatically
while today's analysis software often requires dedicated configuration
and tuning for each data-taking period. Belle II have taken some steps
to address this by using the ``global tag'' concept commonly used in
reconstruction and applying it to analysis, allowing users to better
organise and store their analysis metadata. Nevertheless, until all
metadata is organised under the same global tag umbrella and accessed
through a coherent interface, which is extremely challenging, the
problem remains. Improvements here are necessary both to permit
efficient studies on partial datasets, and to reduce the risk of user
error in metadata access.

The growing complexity of analysis codes makes them extremely fragile
(i.e. bug-prone), hardly reusable, and unsuited for analysis
preservation needs. The current best efforts at analysis preservation
are based on a snapshot of the full analysis setup that can ensure the
possibility to re-run the code on the original dataset in a few years.
This may allow a future analyser to reproduce previous results, but will
likely not provide any understanding, for the future analyser, about
what the analysis was effectively doing. Another use case is to reuse
the analysis code on new data, be that the same dataset with improved
calibrations or additional data that will improve the precision of the
measurement. This brings significant additional challenges, not least
related to metadata.

Nevertheless, these preservation efforts encourage better organisation
of the analysis as a workflow, and promote the use of version control
tools together with continuous integration, which offers a natural route
to improving analysis code quality. In addition to these code quality
measures, a longer-term solution for code complexity may be the adoption
of a Domain-Specific Language (DSL) \cite{Ref10}, discussed in more detail
later, a model that would have physicists write logical descriptions of
their analyses, leaving low-level code generation and hardware
optimisation to a backend. Thus analysis design could be quickly
understood and shared, sidestepping the usual dearth of documentation,
while simultaneously isolating analysis design from implementation and
hardware, providing a natural means of analysis preservation. An
interesting development in this direction is the REANA platform \cite{Ref21}
that supports multiple backends and uses a DSL to provide a framework to
run an analysis. Already a valuable resource as a tool for analysing
open data, it is interesting to ask what would happen if every analyst
first had to structure their analysis and capture their workflows before
they started submitting jobs.

\hypertarget{plans-and-recent-progress}{%
\subsection{Plans and Recent Progress}\label{plans-and-recent-progress}}

Most of the problems identified above were already identified in the
community white paper \cite{Ref17} from 2017 and since then some progress
was made prototyping new technologies to enable data analysis at the
HL-LHC.

A first branch of new technologies is that of efficient data analysis
processing in the last steps of the analysis, i.e. when event data needs
to be selected, derived and aggregated into histograms. Several
platforms developed by industry or data science have emerged in recent
years to quickly aggregate large datasets with parallel execution
(either on clusters of computers or simply on multi-core servers). Many
of those tools have been tested to understand the feasibility of usage
in the HEP context providing fresh ideas and new paradigms that have
stimulated development within the HEP toolkit.

In particular, the Python ecosystem, implementing many such solutions
(e.g. numpy, pandas dataframes), is emerging as a possible alternative
to HEP's traditional C++ environment. This is thanks largely to a
combination of its versatility as a language, allowing rapid
prototyping, and the ability to out-source compute-intensive work to
more performant languages like C and C++. This is similar to the
experimental software frameworks where Python is used as steering code
and C/C++ is used as an efficient backend implementation. The typical
Belle II analysis starts with Python using a custom DSL to specify
particle selections and algorithms implemented in C++ and define output
ntuples. The ntuples are typically analysed by the PyHEP toolkit
\cite{Ref6}, and packages from the wider data science community. Deeper
integration with external Python tools is also particularly important
for enabling state-of-the-art machine learning frameworks such as
PyTorch and TensorFlow, whose relevance in analysis is likely to grow in
the long term. Meanwhile PyROOT, which allows the use of any C++ class
from Python, has the potential to provide a coherent analysis ecosystem
with an effortless integration of HEP specific tools written in C++ with
industry Python big data toolkits.

The ROOT package, which is the foundation and workhorse for LHC analysis,
has been the focus of substantial development. On the data storage
front, the ROOT team demonstrated better data compression and
accelerated reading for a wide variety of data formats. Information
density drives compression and this can vary massively between
experiments and analyses. This new RNTuple format \cite{Ref9, ROOT-2020-HL-LHC}, an
evolution of the TTree file format, shows robust and significant
performance improvements ($1-5\times$ faster), which could potentially save
significant storage space ($10-20\%$) for the HL-LHC. Another area of
progress, and potential consolidation in ROOT I/O, is lossy compression,
where the number of significant bits for a quantity may be far fewer
than a standard storage type's mantissa, enabling further savings in
storage space~\cite{ROOT-2020-HL-LHC}.

Inspired by data science tools, the event processing framework was
extended with DataFrame-like functionality (RDataFrame) implementing a
declarative approach to analysis processing in either Python or C++
(but always executed in C++), which natively exploits multi-core
architectures \cite{Ref7} and could be extended to accelerators. Used from
Python, it allows pre-processing data in C++, and exporting to numpy.
Flexibility in interfacing RDataFrame to non-ROOT data formats has
allowed the ALICE collaboration to prototype its Run-3 analysis model on
this new technology. Growing use of Machine Learning was also
anticipated, and the TMVA toolkit has been improved with substantial
optimisations, more complex deep learning models and enhanced
interoperability with common data science tools \cite{Ref14,Ref15}, which is of
particular importance for inference.

A complementary approach provided in Python is the toolchain of uproot,
awkward arrays and coffea \cite{Ref8} for efficient columnar operations and
transformations of the data. These function as a lightweight end-stage
analysis environment that provides only the elements necessary for
pre-processing ROOT inputs and converting them into formats used by the
standard ML frameworks, typically numpy. Lightweight distribution using
package managers such as pip or conda allows for rapid setup and
extension with other Python tools. Good performance appears to have been
achieved within the scope of these projects, which has been kept
focused, but for the purposes of fair comparison with other existing or
emerging options, defining clear benchmarks for I/O and processing speed
is essential.

A further feature linked to the growing use of Python is the use of
``notebook'' technology such as Jupyter. This permits quasi-interactive
exploration where the annotated history, including formatted outputs and
graphics, can be saved and shared with collaborators for reproduction and
adaptation. Although not a substitute for command line scripts in
well-tested and complex workflows, notebooks make an effective vessel
for software education and have been leveraged as a powerful frontend
for access to facilities including CERN's SWAN \cite{Ref20}.

\hypertarget{prospects-and-rd-needs}{%
\subsection{Prospects and R\&D needs}\label{prospects-and-rd-needs}}

As the complexity of analysis data handling and processing grows,
developing efficient and robust code using a steadily increasing number
of tools running on ever more heterogeneous hardware becomes more and
more difficult. In addition, analysis code is typically re-implemented by
tens of individuals, in different experiments, analysis teams or
institutes mostly performing the same kind of operations, i.e. data
reduction, plotting, variation of systematic uncertainty, fitting, etc.
It is clear that tools and frameworks for performing these repeated
operations need to provide very efficient interfaces to avoid analysis
code bloat. Declarative rather than imperative programming is becoming
visibly more prevalent both in and outside of the field as it naturally
provides efficient syntax.

DSLs are a generalisation of this declarative concept and have already
demonstrated their ability to simplify analysis code. Not only do 
they allow users to express operations concisely, improving
comprehension and reusability, DSLs provide a natural layer of
insulation against hardware evolution as low-level, optimised code
generation for multiple different platforms and architectures is
delegated to the tool and framework experts. Prototypes of DSLs have
been developed in recent years \cite{Ref10} for the event processing and
histogram production parts of the analysis workflow, and even built into
experimental frameworks to perform high-level analysis \cite{Ref19}. Similar
efforts can also be investigated in the context of data interpretation
and statistical analysis and an interesting example here is the so
called ``combine tools'' developed by CMS and based on RooFit for the
Higgs discovery. Descriptions of a full analysis in terms of 
workflows could then tie together the analysis at the top level \cite{Ref18,Ref21}. With some effort, the HEP community may be able to converge on
effective tools that provide a common solution, ideally across
experiments.

There is an obvious need for good integration with analysis backends,
whether these be local CPU, batch or grid resources. At some stage in
the analysis process, interactivity or fast turn-around is needed. While
typical grid task completion has a time scale of one day to one week,
fast turn-around exploration requires answers within a few seconds up to
perhaps a few hours in order to keep people productive. As analysis
tasks are often I/O-intensive and not necessarily-CPU intensive, the
right trade off between CPU and high bandwidth storage should be
carefully studied. More detailed studies of resource usage and its
evolution, especially as pertains to the need for heterogeneous compute,
is needed to ensure the right resource balance is available in the
HL-LHC computing infrastructure. This will require more interaction
between the analysis and infrastructure communities.

As a proof of principle, an unoptimised Higgs discovery analysis has
been rerun on Google Cloud in a few minutes as the infrastructure was
able to quickly provide tens of thousands of CPUs with guaranteed
bandwidth to storage of 2 Gb/coresec \cite{Ref12}. Scaling this to
sustained, diverse analysis payloads will be far from trivial, so the
question becomes how we can compare different approaches to provisioning
analysis. Eight initial benchmark challenges have been defined to set
the scope that should be addressed by analysis systems \cite{Ref11}. What
remains to be demonstrated is the capability of such systems to scale to
full analysis complexity, including tasks such as the handling of
systematic uncertainties and data-driven background estimation, and
these are yet to be incorporated into the benchmarks.

In contrast to the activity on analysis languages and infrastructure,
metadata is a topic that is almost entirely neglected by the analysis
community, even while computing experts have long understood its
importance. That is likely due to the absence of a single coherent
stake-holder for and provider of analysis metadata. Event data comes in
a ROOT file, but already within a single experiment there are numerous
sources of metadata that operate on vastly different timescales and this
poses a significant challenge to standardisation. As analyses and
detectors grow in complexity, further divergence may amplify the
challenge, making it critical to begin addressing this problem promptly.
Nevertheless, the drive to capture a complete analysis must drag
metadata with it, and the gleeful adoption of the declarative paradigm
will make database-like interfaces seem less scary than they once were
and may finally allow a more complete integration with analysis code. 

One final comment in reviewing recent developments is worth noting.
Exposure of the HEP community to industry-standard data science tools
has undoubtedly been a force for good and a net gain for the community,
providing fresh ideas and new ways of working - RDataFrame is an
excellent example. It is clear that the larger and better-supported data
science community will continue to innovate and produce tools that HEP
will benefit from, both by using directly and by borrowing their ideas
for our own tools. Taking machine learning as an example, it is
extremely likely that industry will blaze a trail that HEP would do well
to follow and we should continue to build bridges to those tools.
Equally though, HEP will continue to have its own unique challenges and
it is less likely that ultra-performant machine learning inference will
be top of the data scientists wishlist. Given the challenges presented
by extreme-throughput analysis at the HL-LHC, the right balance must be
found between software development within the community and relying on
external toolkits.

Finally, attention must be given to ensuring the longevity of any
projects that the HEP community decides to adopt en masse. Whether by
specific institutes, funding schemes or other collective programmes,
responsibility for key software will need to be ensured in the long
term. The development of a secure support base will therefore need to be
considered for emerging projects that show significant promise.

\hypertarget{summary}{%
\section{Summary}\label{summary}}

The main challenges and R\&D lines of interest have been discussed
above. There are clearly identified priority topics for the different
areas in the next few years\footnote{Longer term R\&D, into high-risk, high-reward blue sky areas,
such as quantum or neuromorphic computing, should happen at a low level,
but are almost certainly too far from production to deliver improvements
for Run 4 of the LHC.}.

\hypertarget{physics-event-generators-1}{%
\subsection{Physics Event Generators}\label{physics-event-generators-1}}

\begin{enumerate}
\def\labelenumi{\arabic{enumi}.}
\item
  \begin{quote}
  Gain a more detailed understanding of current CPU costs by accounting
  and profiling.
  \end{quote}
\item
  \begin{quote}
  Survey generator codes to understand the best way to move to GPUs and
  vectorised code, and prototype the port of the software to GPUs using
  data-parallel paradigms.
  \end{quote}
\item
  \begin{quote}
  Support efforts to optimise phase space sampling and integration
  algorithms, including the use of Machine Learning techniques such as
  neural networks.
  \end{quote}
\item
  \begin{quote}
  Promote research on how to reduce the cost associated with negative
  weight events, using new theoretical or experimental approaches.
  \end{quote}
\item
  \begin{quote}
  Promote collaboration, training, funding and career opportunities in
  the generator area.
  \end{quote}
\end{enumerate}

\hypertarget{detector-simulation-1}{%
\subsection{Detector Simulation}\label{detector-simulation-1}}

\begin{enumerate}
\def\labelenumi{\arabic{enumi}.}
\item
  \begin{quote}
  Undertake a detailed performance analysis of current production
  detector simulation code, to better understand where performance
  limitations in data and cache locality arise from.
  \end{quote}
\item
  \begin{quote}
  Improve current codes through refactoring and specialised HEP models
  to avoid bottlenecks identified.
  \end{quote}
\item
  \begin{quote}
  Develop further fast simulation models and techniques, including
  machine learning based options in addition to optimised parametric
  approaches as well as common frameworks for their tuning and
  validation.
  \end{quote}
\item
  \begin{quote}
  Undertake R\&D into GPU codes for tackling very specific, time consuming, parts of
  the simulation for HEP. Specifically calorimetry, including geometry,
  field and electro-magnetic physics processes.
  \end{quote}
\item
  \begin{quote}
  Develop integration prototypes to exercise and benchmark the
  simultaneous use of GPU specific libraries with CPU-based software
  (e.g. Geant4) to cover all particles and processes required in
  experiments simulation frameworks.
  \end{quote}
\end{enumerate}

\hypertarget{reconstruction-and-software-triggers-1}{%
\subsection{Reconstruction and Software
Triggers}\label{reconstruction-and-software-triggers-1}}

\begin{enumerate}
\def\labelenumi{\arabic{enumi}.}
\item
  \begin{quote}
  Develop fast and performant reconstruction software, making optimal
  use of hardware resources. A particularly crucial development for
  HL-LHC is that of pattern recognition solutions (including timing
  information, if available) that can withstand the increase of pile-up
  at HL-LHC. Wherever possible, solutions should be uniform for online
  and offline software, so the experiments can benefit from consistent
  event selection between trigger and final analysis and from more
  effective real-time analysis results.
  \end{quote}
\item
  \begin{quote}
  Develop solutions that can exploit accelerators (FPGA, GPU) for
  trigger and reconstruction, including general-purpose interoperability
  libraries with a focus to generalising and mitigating their
  dependencies on architect\-ure specific codes.
  \end{quote}
\item
  \begin{quote}
  Further promote activities in the area of data quality and software
  monitoring, as the quality of the recorded and reconstructed data is
  paramount for all physics analysis.
  \end{quote}
\item
  \begin{quote}
  Support efforts towards the implementation of machine learning models
  in experimental framework (C++) to enable more widespread use in
  trigger and reconstruction software.
  \end{quote}
\end{enumerate}

\hypertarget{data-analysis-1}{%
\subsection{Data Analysis}\label{data-analysis-1}}

\begin{enumerate}
\def\labelenumi{\arabic{enumi}.}
\item
  \begin{quote}
  Identify a CPU-efficient data reduction and storage model that serves
  the majority of analyses with a footprint of O(1kb/event), retaining
  the flexibility to accommodate the needs of specialised analyses.
  \end{quote}
\item
  \begin{quote}
  Streamline analysis metadata access and calibration schemes, and
  provide effective book-keeping for fractional datasets.
  \end{quote}
\item
  \begin{quote}
  Develop cross-experimental tools establishing declarative syntax
  and/or languages for analysis description, interfaced with distributed
  computing backends.
  \end{quote}
\item
  \begin{quote}
  Define and improve schemes for interoperability of end-stage
  experimental software with data science and machine learning frameworks.
  \end{quote}
\end{enumerate}

Undertaking this programme of priority topics and R\&D requires, first
and foremost, investment from funding agencies to support developments.
A new generation of developers is also needed to refresh and regenerate
the long lived software projects on which the field so heavily relies.
The transition from R\&D to production ready software is usually long
and so it is urgent to undertake this investment now, with the positive
outcomes ready to become integrated in time for HL-LHC. Regenerating the
cadre of software developers in HEP also requires support for realistic
long term career prospects for those who specialise in this vital area.
Collaboration with industry can certainly be fruitful and is to be
encouraged to allow early access to technology and communication of HEP
problems and priorities.
R\&D that is successful will become the next generation of production
software, but this will require lifetime support for maintainance and
further evolution.

Almost all the domain areas identify the use of compute accelerators,
particularly GPUs, as being a priority item, which matches our
expectation of how computing hardware will evolve. As a common problem
area, this is an obvious area in which expertise should be developed.

The stakeholders in HEP must be able to give their priorities and
feedback on R\&D areas. Each project has mechanisms for doing this,
while the HSF can help to provide overall input on prioritisation
and coordination of common cross-experiment activities.

\sloppy
\raggedright
\clearpage
\printbibliography[title={References},heading=bibintoc]

\end{document}